\documentclass{aa}
\usepackage{epsfig}
\usepackage{amsmath}
\usepackage{graphicx}
\usepackage{longtable}
\usepackage{rotating}
\usepackage{natbib}
\bibpunct{(}{)}{;}{a}{}{,} 

\newcommand{\bv}{$(B-V)$}
\newcommand{\mv}{$M_V$}

\newcommand{\mh}{$\left[\frac{M}{H}\right]$}
\newcommand{\logg}{$\log g$}
\newcommand{\teff}{$T_{\rm eff}$}
\newcommand{\grad}{$^\circ$}

\newcommand{\delpi}{$\frac{\Delta \pi}{\pi}$}
\newcommand{\mua}{$\mu^*_{\alpha}$}
\newcommand{\mud}{$\mu_{\delta}$}
\newcommand{\kms}{${\rm km\ s^{-1}}$}

\newcommand{\vrad}{$v_{\rm rad}$}
\newcommand{\vpec}{$v_{\rm pec}$}
\newcommand{\msun}{M$_{\odot}$}
\newcommand{\hip}{{\sc Hipparcos}}
\begin{document}

\authorrunning{Kaempf et al.}
\titlerunning{Kinematics of RHBs}
\title{Kinematics of RHB stars to trace the structure of the galaxy}

\author{T.A. Kaempf\inst{1} \and K.S. de Boer\inst{1} \and  M. Altmann\inst{1,2,3}}

\offprints{{\tt tkaempf@astro.uni-bonn.de}}

\institute{ Sternwarte der Universit\"at Bonn, Auf dem H\"ugel 71, D-53121 Bonn,
Germany
\and Dr. Remeis Sternwarte der Universit\"at Erlangen-N\"urnberg, Sternwartstr.
7, D-96049 Bamberg, Germany
\and Departamento de Astronom\'{i}a de la Universidad de Chile, Camino del
Observatorio 1515, Las Condes, Chile}

\date{Received 16 November 2004 / Accepted 16 November 2004}

\abstract{Red horizontal-branch (RHB) stars have been selected from the \hip\
catalogue to investigate their kinematics and spatial distribution.
\hip\ parallaxes, literature radial velocities and
\hip\ proper motions,
together with models for the gravitational potential of the Milky Way allow
a calculation of the actual velocity vectors and the orbits of the RHB stars.
The velocity characteristics are used to define a halo population sample (HPS)
in the collection of RHBs.
The orbits lead statistically to an overall $z$-distance probability
distribution, showing that the RHBs exhibit two populations,
a disk one having a scale height of $h_{\rm disk} \simeq 0.6$~kpc
and a halo one of $\simeq 4$~kpc.
We have investigated the influence on our results
of parallax accuracy and of a demarcation line
in the HRD between the RHB and the red-giant (RG) star region.
Neither of them show marked effects.
We have performed the orbit analysis using the potential model
of Allen \& Santillan as well as of Dehnen \& Binney.
The results differ only slightly for the disk population, showing that these
potential models are not a critical part of such orbit investigations.
RHB scale height values are smaller than those found earlier for sdB stars,
most likely because the samples of stars used had different spatial
distributions a priori.
The data do not allow us to specify a trend in the kinematic behaviour of
star types along the horizontal branch.

\keywords{astrometry -- Stars: kinematics -- Stars: horizontal branch -- Galaxy:
general}} 

\maketitle


\section{Introduction}
\label{intro}

While it is rather simple to study the structure of other stellar systems from
images alone, the true shape of the Milky Way is very hard to determine.
With the help of ever growing galactic surveys one is finally able to
make use of highly accurate positions and motions of thousands of stars.
In conjunction with mass models of the Milky Way, such a wealth of data gives
us the first detailed glimpses of the structure and evolution of our own galaxy.

Choosing a single stellar type, in particular one with well-defined
luminosity, allows us to work with clean samples.
Late main-sequence (MS) stars were used to probe the solar vicinity within
several hundred parsecs, while giants and A type MS stars were observed up to
two kiloparsecs away.
For regions of negligible interstellar extinction the stellar density field
was derived from these surveys (e.g. \citealt{Becker40}, \citealt{Elv51},
and \citealt{Up63}).
Results from these studies established the idea of a multi-component galaxy,
consisting of a central bulge, a thin disk, a thick disk (found by
\citealt{GR83}) and an encompassing halo.

Further information on the exact shapes of these components can then be used to
constrain theories of galactic evolution.
Models for the formation history of the disk are separated into two
families.
If the thin disk is formed after the thick disk, one speaks of a
``top down'' scenario.
Different examples of this model type are described in 
\cite{Gil84}, \cite{San90}, \cite{NR91}, and \cite{Bur92}.
If the thin disk is formed before the thick disk, it is called a
``bottom up'' scenario.
Members of this group are the models of \cite{Nor87}, \cite{Qui93}, \cite{K2k2}
and \cite{GR83}.
A summary of the various models is given by \cite{Maj93}.

With the astrometry satellite \hip, an
automated survey of stellar positions with unprecedented accuracy was made
possible.
Some 120000 stars were observed for the main mission.
Data were collected in the Hipparcos Catalogue (\citealt{HIP}).
In addition, the satellite provided two-colour photometry for slightly more than
one million stars.
The data were reported in the Tycho Catalogue
(\citealt{HIP}).

Among the brighter old stars easily identifiable from optical observations
are the RR~Lyr stars and blue sub\-dwarf stars. 
These are stars of the horizontal branch (HB), 
consisting of a $\simeq 0.5$~\msun\ He core surrounded by a hydrogen shell 
ranging in mass from negligible to up to about 0.5~\msun. 
From thin to thick shells, the stars are classified as 
sdO, sdB, HBB and HBA stars, followed by the RR~Lyr stars, 
and finally the ``massive'', red HB (RHB) stars. 
Of these, the sdB and RR~Lyr stars are the easiest to spot and characterize.
The former show a simple relation between the line strengths of the Balmer
series and \teff\ and \logg.
The latter, as variable stars, are identified by their characteristic light
curve.

Several studies exist dealing with the spatial distribution of sdB stars. 
\cite{Heber86} and \cite{Moehler90} found vertical distributions 
with scale heights of $\simeq 200$~pc, way too small for old stars. 
However, 
\cite{Villeneuve95a} and \cite{Villeneuve95b} showed that larger values are
more realistic. 
In two extensive studies of the kinematics of sdB stars 
(\citealt{B97}, \citealt{AB2k4}) 
it could be shown that the sdB stars seen in the solar neighbourhood 
are divided in two populations, 
a disk one with a scale height of $\simeq 1$~kpc and an extended one 
chracterized by a scale height of $\simeq 7$~kpc. 
Furthermore, trends were seen in the kinematics of HB star samples. 
The cooler Blue HB (BHB) star samples seemed to be more
dominated by the halo population than the sdB star sample (\citealt{AB2000}).

\begin{figure}
\centering
\epsfig{file=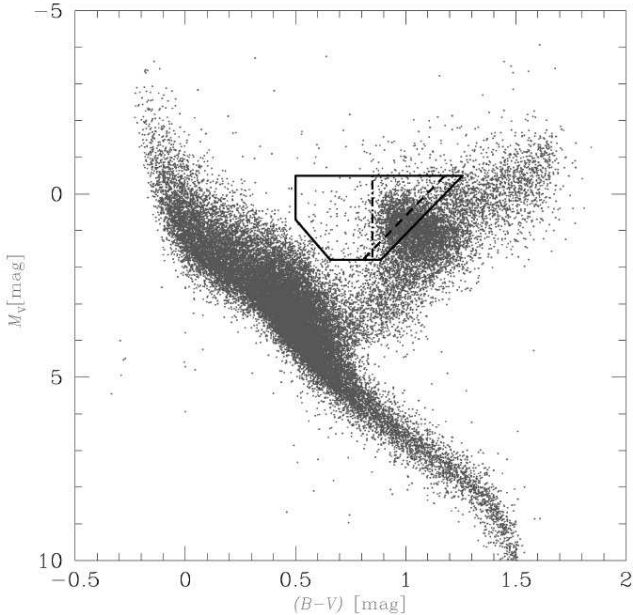,scale=0.4, angle=0}
\caption{CMD of single \hip\ stars with $\Delta(B-V)\leq0.025$ and
$\frac{\Delta \pi}{\pi}\leq20\%$. The stars inside the window are our selected
RHB candidates. The window borders are: \mbox{$(B-V)\geq0.5$},
$M_V\in\{-0.5\,1.8\}$, $M_V\leq-2.8+7\cdot(B-V)$,
\mbox{$M_V\leq7.3-6.2\cdot(B-V)$}.
The dash-dotted line marks the window separation border of
Sect.\,\ref{bluered} while the dashed line marks the alternative cut-off of
Sect.\,\ref{res5}.\label{CMDsel}}
\end{figure}

In this study we investigate the kinematics and spatial distribution 
of the red HB stars. 
Finding RHBs in a sky field is not without difficulty.
By just looking at colours, one can easily confuse them with MS
stars, subgiants, red-clump stars and blue-loop stars.
Without knowledge of absolute magnitudes it is therefore not possible to select
RHBs just by colour.
Using \hip\ data we are able to resolve this matter.
Moreover, it gives us a rather large sample of RHBs to work with.

To characterize the {\sl large scale} distribution of stars 
of a given type one in principle would need to know the location 
of such stars in complete large distance samples.
For our {\sc Hipparcos} RHB stars this is illusionary, 
since the sample is complete out to perhaps 100 pc 
(see Sect.\,\ref{overall}). 
However, the kinematics of the RHB stars also give information 
about past and future locations. 
We will calculate the galactic orbits (as in \citealt{B97}, \citealt{AB2k4})
of the RHB stars and analyze the orbit statistics to arrive at mean spatial
distributions. 

The organisation of this paper is the following.
We begin with the assembly of our data sample in Sect.\,\ref{sample}.
Errors and selection effects of the sample are described in Sect.\,\ref{from}.
The analysis of the orbits of the selected RHBs is given in Sect.\,\ref{result},
while Sect.\,\ref{models} compares results using different models for the
galactic potential.
We conclude with a final discussion in Sect.\,\ref{final}.


\section{Establishing the RHB data sample}
\label{sample}

\cite{Taut96} and \cite{rose85} give observational methods to find RHB stars
through photometriy and spectroscopy.
However, the simplest way to define the state of evolution of stars is given by
their location in the HRD.

Using \hip\ data we created a colour magnitude diagram of stars not listed in
the {\it Catalogue of the Components of Double
and Multiple Stars} (\citealt{CCDM}), having
\bv$\leq 0.025$~mag, and relative parallax error smaller than $20$\%.
The RHB stars could then be selected from the diagram by choosing an appropriate
range in $M_V$ and \bv\ by eye (see Fig.\,\ref{CMDsel}).
CMDs of globular clusters like 47 Tuc helped choose the proper magnitude and
colour ranges.
We also confirmed that the spectroscopically identified RHB stars
from \cite{Behr03} mostly fulfil the selection criteria.

\begin{table}
\centering
\caption{Numbers of stars at various selection steps as a function of relative
parallax error. The three BB2k columns show: number of RHB sample stars found,
total number of measurements, number of stars that are fainter than 7.3
mag, the completeness limit of the \hip\ mission. \label{startab}}
\begin{tabular}{cccccc}
\hline
\noalign{\smallskip}
\delpi & HIP & RHBs & \multicolumn{3}{c}{BB2k} \\
& & & stars & meas. & $>$7.3 mag \\
\hline
$\leq10$\% & 17295 & 547 & 455 & 480 & 0 \\
$\leq15$\% & 28700 & 1393 & 828 & 890 & 28 \\
$\leq20$\% & 38114 & 2241 & 1044 & 1127 & 103 \\
$\leq25$\% & 45622 & 3084 & 1215 & 1313 & 211 \\
$\leq30$\% & 51231 & 3795 & 1327 & 1437 & 305 \\
\hline
\end{tabular}
\end{table}

For more on the complexities of the boundary line toward the giant branch 
see Sect.\,\ref{RHBRC}.

Five samples were assembled from stars within this window.
These sets differ in parallax precisions, with each having \delpi\ lower than
10\%, 15\%, 20\%, 25\%, and 30\%, respectively.
Thus, with such a criterion, we almost get a volume-complete sample for
spheres with a rather blurry boundary.
Effects from this sample definition are discussed in Sect.\,\ref{res4} for the
10\%, 20\%, and 30\% samples.
Table \ref{startab} gives the number of stars in the samples at different
points of the assembly process.

To reconstruct the full spatial motion, radial velocities are required.
\hip\ did not have a dedicated instrument to measure \vrad\ and so
we chose the recent release of the {\it General Catalog of mean radial
velocities} by \cite{bb}; (hereafter BB2k).
By cross-checking our candidate RHBs with BB2k we found radial velocities for
35\%\ up to 83\%\ of our HIP stars, depending on the parallax accuracy
(Table\,\ref{startab}).
This already hints at selection effects that are discussed in
Sect.\,\ref{select}.


\section{Error limits and selection effects}
\label{from}

\subsection{Error limits}
\label{err}

Measurement errors for the \hip\ objects vary not only with brightness but also
with celestial position.
This is due to the scanning law because certain areas in the sky were
observed more often than others.
Therefore, for all HIP stars of $Hp<9$~mag (the faintest of our
stars has $Hp=8.86$~mag) we only cite median precisions.
They are 0.97~mas for the parallaxes and 0.88~mas/yr and 0.74~mas/yr for
\mua$=\mu_{\alpha}\cdot \cos\,\delta$
and \mud, respectively.
Systematic errors were estimated to be below 0.1~mas.

Errors from BB2k depend on the source from which the radial velocity was taken.
Single high resolution spectra, of course, yield better results compared to
big surveys at low resolution.
However, errors are usually below 10~\kms\ and never higher than
20~\kms\ for our samples.

\hip\ proper motion uncertainties for stars of our sample are equal to $\simeq
2$~\kms.
Therefore the total error in
space velocity is dominated by the uncertainty of the radial component.
Thus, we estimate the total error in space velocity to be about
10~\kms.


\subsection{Possible selection effects}
\label{select}

\subsubsection{\hip\ stars and measurement accuracy}
\label{hipacc}
\hip\ was an astrometric mission that used a list of objects to be
observed, the \hip\ Input Catalogue (HIC) by \cite{hic}.
A magnitude-complete set down to 7.3 mag was assembled, some 52000 objects.
Any stars fainter than that limit came from individual proposals.
The selection effects so introduced cannot be quantified. 
For the number of stars in our sample fainter than 7.3 mag 
see Table \ref{startab}. 

\hip\ parallaxes are usually determined at an average precision of 0.8~mas for
high ecliptic latitudes, and around 1.2~mas near the ecliptic.
High accuracy areas for \mua\ are roughly at $|\beta| \ge 40$\grad
(0.8~mas\ yr$^{-1}$, down from 1.3~mas\ yr$^{-1}$). 
For \mud\ on average the standard error is 0.8~mas\ yr$^{-1}$.
No position-dependent colour errors are given in \cite{HIP}.

The mission's $Hp$ passband was more sensitive to blue light (Fig 1.3.1 in
\citealt{HIP}).
Redder stars, therefore, show a higher uncertainty in their measurements.
Since our stars are usually brighter than $V=8$~mag, this effect is small and
probably only shows up in the crowded region of the RGB cut-off.

\subsubsection{Boundary of RHB and Red Clump}
\label{RHBRC}

The definition of our RHB sample in the CMD is a selection effect in its 
own right.
While the range in luminosity 
and the boundary on the side of the MS can be easily defined, 
the separation from the red-giant branch (RGB) is less certain. 

In particular, that region of the CMD contains the ``Red Clump'',
a region where three main types of stars are found. 
First, there are the RG stars evolving upwards in the CMD 
(but having an uneven distribution along the RG).
These are mostly contributing to the red part of the Red Clump, and are thus
excluded from our sample.
Second, there are the red HB stars distributed horizontally toward the RGB 
but of which the ones with mass up to 1.5~\msun\ are located above and 
somewhat bluer than the reddest end of the HB range 
(see e.g. \citealt{Sweigart87}, \citealt{Seidel87} for the overall shape of the
HB). 
Finally, the yet more massive blue loop stars may contribute to the
more luminous part of the Red Clump region.
A full discussion of the complexities can be found in \cite{Gallart98}. 

Since all HB stars have essentially the same life time irrespective 
of their mass, the possible presence of younger (and thus more massive) 
RHBs cannot be neglected. 
Young RHBs exhibit disk kinematics and we may expect the disk population 
to be more dominant in our sample. 

Effects of the choice of the boundaries in the CMD 
(see Fig.\,\ref{CMDsel}) will be discussed in Sects.\,\ref{res5} and
\ref{bluered}.

\subsubsection{Radial velocity catalogues}
\label{rvcat}
The BB2k data are a compendium of measured radial velocities from various
sources published before December 1990.
By looking at how many of the RHBs compared to the whole HIP
stars can be found in the BB2k catalogue, we can estimate how well, compared to
other stellar types, our RHB stars are represented in BB2k.

We limit the search to objects brighter than 7.3 mag in $V$ for completeness
reasons.
A total of 15752 single stars fit this criteria, with 2052 falling in
our RHB window.
BB2k lists entries for 9446 (60\%) and 1101 (54\%) of these categories,
respectively.
We conclude that our RHB candidate stars are not significantly undersampled in
BB2k.

\subsubsection{Overall effects}
\label{overall}
Considering the properties of the catalogues used and our selection criterion,
we can estimate the volume completeness of our sample.
\hip\ measured all stars down to 7.3~mag.
In conjunction with our lower
absolute magnitude limit of 1.8~mag gives (assuming negligable extinction)
a completeness radius of 126~pc.
The BB2k is not magnitude complete.
Some 1600 RHB candidates with \delpi$\leq$30\% and brighter than 7.3~mag are not
listed in BB2k.
A merging of the two catalogues thus shrinks the completeness radius.
For \delpi$\leq$30\% the new limiting magnitude is 6.4~mag corresponding to a
distance of 83~pc.
In contrast to \cite{AB2k4}, we are
not limited towards the bright magnitudes.
Thereby it is certain that all nearby candidate RHBs are included in our sample.

\begin{figure}
\centering
\epsfig{file=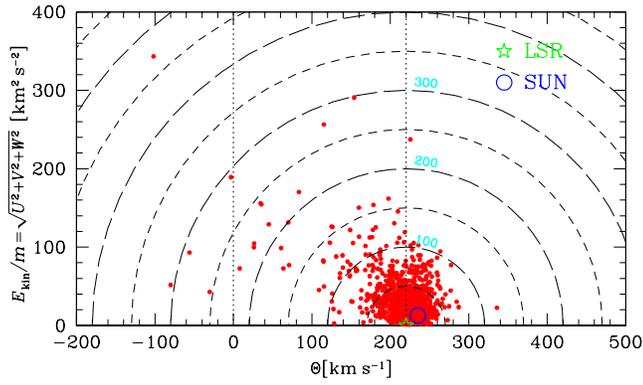,scale=0.43, angle=0}
\caption{The Toomre diagram of our \delpi $\leq30$\%\ sample. The circles show
the peculiar velocity $v_{\rm pec}=\sqrt{\Phi^2+W^2+(\Theta-\Theta_{LSR})^2}$
with respect to the LSR. Up to a peculiar velocity of 50~\kms\ the sample is
distributed rather evenly. Upwards of these values the stars fall short in
orbital velocity. These objects are halo and thick disk stars.
\label{Toomre}}
\end{figure}

For all objects with $V\leq7.3$ mag we know the HIC to be complete.
Measurement errors of these objects are such that we can confirm the correct
CMD position for 91.3\%\ of our RHB sample.
The only other factor of influence is the BB2k, which gives $v_{\rm rad}$
for $\sim$54\%\ of the selected HIP stars.
A total percentage of completeness for these bright stars is then found to
be 91.3\%$\cdot$54\%$\simeq$50\%.
For our fainter objects this value decreases to $\sim$20\%\
if we assume that the HIC only contains half of all fainter RHBs.


\section{The spatial distribution of RHB stars}
\label{result}

\subsection{Distribution from position and velocity}
\label{res1}

A histogram of the circular velocity $\Theta$ of our sample shows
a high narrow peak located at roughly $\Theta_{\rm LSR}$.
These are the thin disk stars.
After taking out this group,
the presence of a broader distribution centered around $\Theta \simeq 180$~\kms\
becomes apparent. 
It may represent the group often called the thick disk.
Finally, there are stars covering a wide range of values of $\Theta \in
[-100,+340]$~\kms.
These are the halo stars in our sample.
However, we cannot assign an individual star to these groups 
and therefore no velocity dispersions were calculated.

\begin{figure}
\centering
\epsfig{file=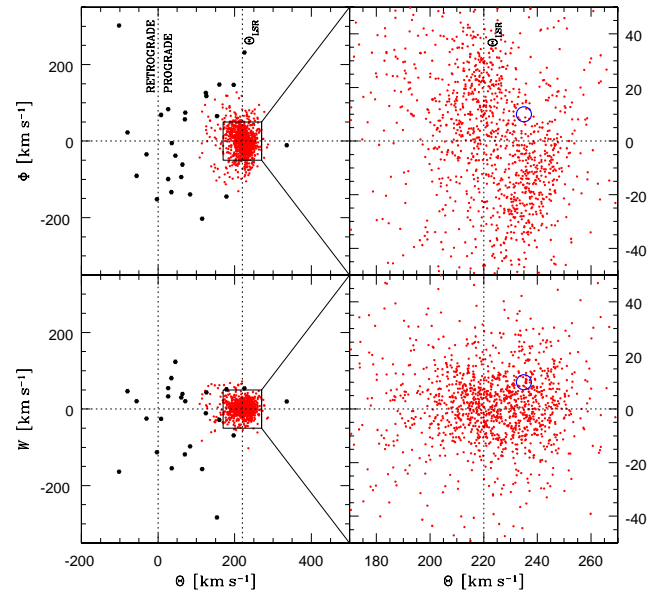,scale=0.42, angle=0}
\caption{The Bottlinger diagram of our \delpi$\leq30$\%\ sample. The
panels at right show the enlarged regions marked by squares in the left panels.
Heavier points are for the Halo Population Sample (HPS) stars.
The circle symbols denote solar values.
Note the two groups of stars in the more detailed $\Theta,\Phi$ panel.
\label{bott}}
\end{figure}

The separation of the thick and
thin disk stars can be seen clearly in a Toomre diagram
(Fig.\,\ref{Toomre}).
A large number of objects cluster in the region with \vpec $\leq50$~\kms.
These are mainly thin disk stars that are moving along with the LSR.
Stars in a range of $50 \leq$\vpec$\leq150$~\kms\ are mostly thick disk objects.
They are lagging behind in $\Theta$ because they are kinematically heated up
compared to thin disk stars.
This effect is known as the asymmetric drift of the thick disk (see e.g.
\citealt{BT}).

All other stars, upwards of \vpec$=150$~\kms, must be halo objects.
Some have large perpendicular velocities, both with prograde and retrograde
orbits.
Stars with large peculiar velocities include those staying in the disk but
having highly non-circular orbits.
An example of these is the star with $\Theta \simeq 340$~\kms, but which
never strays further than 530~pc from the galactic plane.
We make a rough selection of halo objects for further analyses and will refer
to these 25 stars as the Halo Population Sample ({\bf HPS}).

The separation of the populations can be further investigated with the help of
a Bottlinger diagram (Fig.\,\ref{bott}).
Stars of the above defined HPS are marked by heavier points and show
velocities that are very different from the LSR.
Other stars outside of the enlarged frames (Fig.\,\ref{bott}, right side) are
most certainly thick disk stars.

\begin{figure}
\centering
\epsfig{file=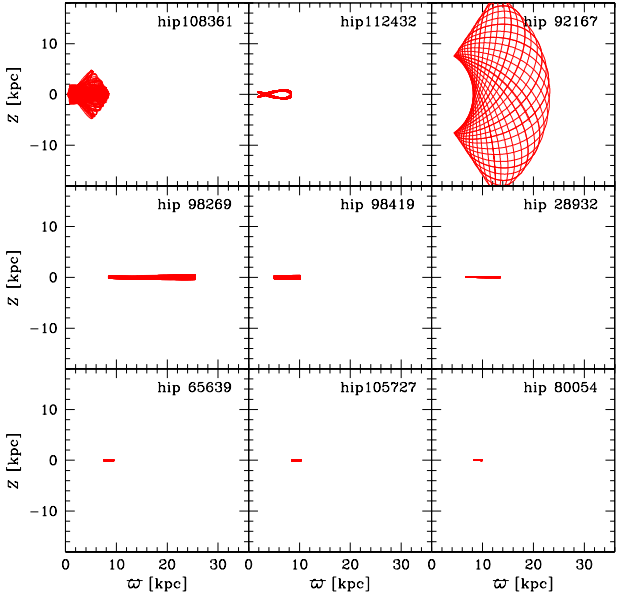,scale=0.43, angle=0}
\caption{Meridian plots of some of the calculated orbits in a model by
\cite{AL91}. The top row shows
orbits of HPS stars. The middle row shows the orbit of the fastrunner star HIP
98269 (left) while the other two are thick disk candidates. The
bottom row depicts thin disk orbits.\label{orbs}}
\end{figure}

\begin{figure}[b]
\centering
\epsfig{file=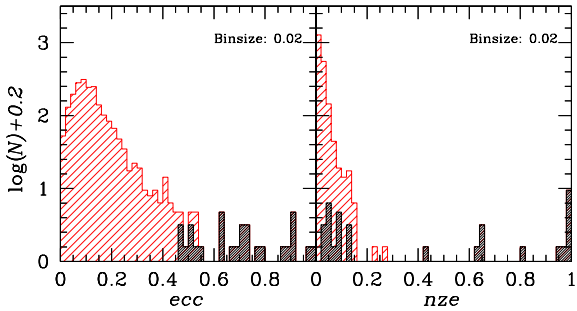,scale=0.4, angle=0}
\caption{Logarithmic histograms, that show the complete \delpi$\leq30$\%\
sample (gray) and the contribution of our HPS stars (black).
We have added a constant offset for visibility reasons.
The $ecc$-histogram shows a separation of populations around 0.45.
As expected, the halo stars show much higher eccentricities than the disk
stars.
In the $nze$-histogram the halo stars do not only build the tail of the
distibution, but also account for low values.
The peak on the right side of the diagram includes the stars with
higher normalized $z$-extent than plotted.
\label{eccnze}}
\end{figure}

The right half of Fig.\,\ref{bott} shows the values similar to the LSR in more
detail.
The $\Theta,\Phi$-diagram shows a separation into two groups.
Such structures are well known and have been analysed in various studies.
Several stellar streams were already found in the $U,V$ distribution of \hip\
data.
\cite{Skuljan99} even associated close star clusters with some of these
streams.
We refer to \cite{Skuljan99} and \cite{Nord04} for more on that topic.

\subsection{Population assignment from orbit data}
\label{res2}

We have calculated the orbits of our stars based on 
the galactic potential model of \cite{AL91} (hereafter AS) 
using an orbit code developed by \cite{Ode92}, and 
further adapted for our work (\citealt{B97}, \citealt{AB2k4}).
Orbits are calculated in steps of 1\,Myr over a time span of 10\,Gyr 
(since we are not modeling the evolution of the Galaxy, this long span,
although physically unrealistic,
is allowed for a good sampling of the shape of wide orbits). 
A selected few orbits can be seen in Fig.\,\ref{orbs}.
Most of the orbits are of a regular box shape.
A few are irregularly shaped, perhaps a result of the close approach of the
star to the galactic center and thus the interaction with the Bulge potential.

\begin{figure}
\centering
\epsfig{file=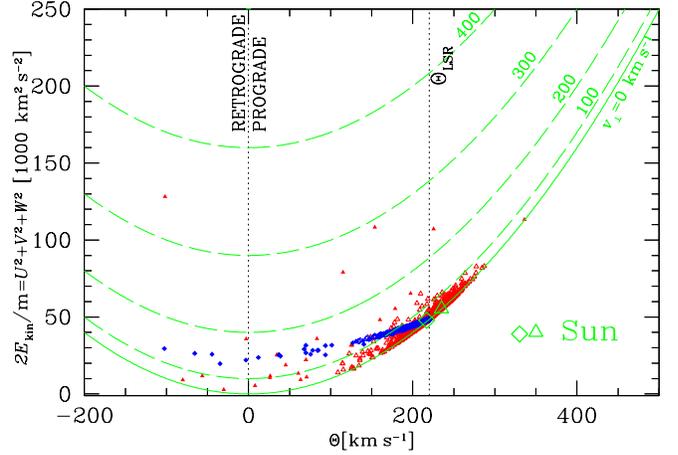,scale=0.45, angle=0}
\caption{Diagram of $\Theta$ (gray triangles) and $\Theta_{\rm med}$ (black
squares) against total kinetic energy and median kinetic energy, respectively.
Filled symbols are stars of the HPS. 
The parabolas are isovelocity lines for $v_{\perp}$, which is
orthogonal to $\Theta$. Medianised values lie on an almost straight line
pointing from the LSR data point towards lower $E_{\rm kin,med}$ and
$\Theta_{\rm med}$.
\label{ekin}}
\end{figure}

By looking at histograms of the orbit eccentricities and normalized $z$-extents
(Fig.\,\ref{eccnze}), we qualitatively come to the same conclusions as in
\cite{AB2k4}.
In the $ecc$-histogram, a separation between a high and low eccentricity
sample can be made with the help of the HPS.
It is roughly at $ecc=0.45$, above which the halo stars dominate the
distribution.
In the $nze$-histogram, the stars at $nze > 0.3$ are clearly of the HPS.
But HPS stars also contribute to the
peak at low normalized $z$-extents, 
since our definition of the HPS does not exclude highly eccentric, coplanar
orbits.

Fig.\,\ref{ekin} shows the kinetic energy versus orbital velocity
of our star sample.
Both the measured and medianised values are given.
Looking at the median values, one can see that most of the disk stars are
situated along a line pointing from the LSR towards lower $E_{\rm kin,med}$ and
$\Theta_{\rm med}$.
A clear gap can be seen at $\Theta\sim110$~\kms, below which there are only HPS
stars.

Using the information from Fig.\,\ref{Toomre} we have separated the HPS stars
from the disk stars, a separation confirmed through the kinematic
representation of Figs.\,\ref{bott}, \ref{eccnze}, and \ref{ekin}.
However, individual stars of one group may still be included in the other group.

\subsection{Spatial distribution from orbit statistics}
\label{res3}

From the sum of the orbits we obtain the $z$-probability distribution of
our sample stars, as introduced in \cite{B97} and further refined in
\cite{AB2k4}.
We analysed it for disk and halo populations with a weighted $\chi^2$-fit.
The large range in $N(z)$ made a weighting scheme necessary that could take
into account both the disk region and halo region data points.
We chose $1/N(z)$ as a weight to accomplish that goal.
A simple exponential relation of the form $N(z)=N(0)\cdot e^{-z/h}$
was fitted for each component.
where $N(0)$ is the star number density in the local galactic plane and $h$
is the scale height.
Both sides of the distribution were used for the calculations.

Analyses of the orbits showed that the data points with $z>8$~kpc are from
three stars only.
They are HD 126778, HD 175305 and HD 184266.
All the other HPS stars show strongly elliptic orbits close to the galactic
plane.

\begin{figure}
\centering
\epsfig{file=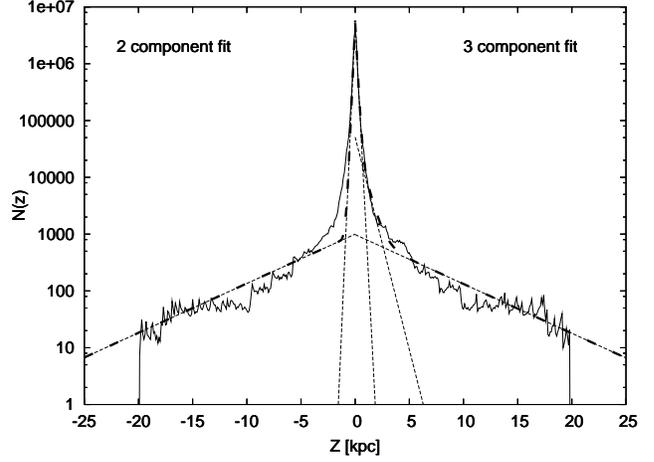,scale=0.35, angle=270}
\caption{The $z$-probability distribution of our \delpi$\leq30$\%\ stars with
a two-component and a three-component $z$-distribution exponential fit,
shown in the left and right halves of the diagram, respectively.
Fitted curves are: thin dashed ones the individual population components (see
Sect.\,\ref{res3}),
thick dashed ones are the complete fits.
Note that the three-component fit is a better approximation of the actual
distribution.
\label{Zhist1}}
\end{figure}

Fig.\,\ref{Zhist1} shows the $z$-probability distribution, as well as the best
fits.
First, a two-component fit, representing the disk and the halo, was made.
But in the range of 2-5~kpc the full $z$-distribution is not very well
approximated (see left half of Fig.\,\ref{Zhist1}).
We therefore also made a three component fit.
The results give a much better fit to the data than with only two components
(right half of Fig.\,\ref{Zhist1}).
Assuming that exponential functions apropriately describe the
$z$-distribution, our data suggests the existence of more than two spatially
distinct stellar groups.

\begin{table}
\caption{The fit results of the $z$-probability distribution from
Fig.\,\ref{Zhist1}. Scale heights and number densities are
given for each fitted component. For a two-component fit no ``thick disk''
values were determined. \label{fits}}
\begin{center}
\begin{tabular}{ccc}
\hline
fit constants& 2 comp. & 3 comp. \\
\hline
$h_{\rm thin}$\,[kpc] & 0.1 & 0.12 \\
$N_{\rm thin}(0)\ ^a$ & 5.5$\cdot10^6$ & 4.8$\cdot10^6$ \\
$h_{\rm thick}$\,[kpc] & -- & 0.58 \\
$N_{\rm thick}(0)\ ^a$ & -- & 51000 \\
$h_{\rm halo}[kpc]\ ^b$ & 5 & 4.2 \\
$N_{\rm halo}(0)\ ^{a,b}$ & 1100 & 900 \\
\hline
\end{tabular}
\end{center}

\noindent $^a$ Values of $N(0)$ come from the numerics of the orbit statistics
and cannot be used as number of stars per unit volume.\\
\noindent $^b$ Halo values are based on three stars only (see Sect.\,\ref{res3}).
\end{table}

The results of the fits are given in Table\,\ref{fits}.
We note that the scale heights of the halo are derived from only three
stellar orbits and are only listed for completeness.

The results obtained are based on what will be called the base-line model. 
Below we will explore star selections with different parallax accuracies 
or selected with different magnitude and colour limits (as in
Fig.\,\ref{CMDsel}).    
We will also explore (in Sect.\,\ref{models}) the effects of different
models for the mass distribution in the Milky Way on our base-line results. 

\subsection{Effects of varying the chosen sample structure}

\subsubsection{Samples with 10,20,30\% parallax accuracy}
\label{res4}

In Sect.\,\ref{sample} we briefly gave some details on star samples with
different \delpi\ constraints.
By analysing the 10\%\ and 20\%\ samples we can derive some information on the
spatial distribution of the stellar populations.
This implies dealing with smaller samples and thus a
deterioration in number statistics.

Diagrams derived from the current velocities are qualitatively the same for all
the samples.
A closer inspection of the Toomre diagram shows that a majority
of our objects are stars with thin disk orbit characteristics and only a minor
fraction is of a halo like type.
Disk stars are included in all the samples in a large
enough quantity for analyses.
In the $U,V$-diagram, the bimodality of the data can still be seen, even in
the 10\%\ sample.
This is due to the fact that the separation is observed for disk stars, and
they are adequately represented in all the percentage sets.
In contrast, only three of the HPS stars have a relative parallax error below
10\%.
This low number implies that with these we cannot make a meaningful
analyses for the halo.
For \delpi$\leq$20\%, the number of HPS stars is twelve.

The orbital data thus confirms the similarity of the composition of the
error-selected samples.
For the disk stars, the histograms are of similar shape
and HPS stars of the 10\%\ and 20\%\ sample are also present with 
eccentricities larger than 0.4.
We conclude that no kinematical preference can be found in the closer samples.

Even though we have very low numbers of stars for a $z$-distribution analysis
in the 10\% sample (see Table\,\ref{startab}), we give rough estimates
on the scale heights.
They are 0.12~kpc for the thin disk, 0.34~kpc for the thick
disk and 8~kpc for the halo.
Scale heights of the 20\% sample are 0.12~kpc, 0.46~kpc, and 9~kpc,
respectively.
These results are not different in an essential way from those of
Table\,\ref{startab}.

\subsubsection{Influences of the RHB--RGB separation line}
\label{res5}

Another, perhaps critical aspect of our RHB selection from \hip\ data is the
heavily populated location of the window border at the RG branch.
By excluding more of the critical red stars, we can check for influences of
giants and see if our above results still hold.

A new separation was adopted, by shifting the line towards brighter \mv\ by
0.5~mag (see dashed line in Fig.\,\ref{CMDsel}).
We thereby excluded almost half of the full \delpi$\leq$30\%\ sample ending
up with some 1942 \hip\ stars.
From these stars, 731 had 789 measurements documented in BB2k (1327 stars in the
full sample, see Tab.\,\ref{startab}).

With this new and restricted sample we noted several things.
The velocity histogram shows that some of the stars lagging
behind the disk rotation are now missing.
The histogram of thick disk velocity stars now consists of several peaks.
Stars with thin disk velocities are also less numerous, but the peak is
still very similar in shape.
Looking at the Toomre diagram allows the identification of the ``missing''
stars.
We found that only 15 of the HPS stars are still included in the
restricted sample,
but the relative number of HPS to disk stars stays the same (1.9\% compared
to 2.1\% for the base-line sample).
In addition, a reduction of star numbers in the range of
100~\kms$\leq v_{\perp}\leq$150~\kms\ was noticed.
The absence of these objects explains new gaps in the
$ecc$-histogram (at $ecc=0.35$) and $\Theta,E_{\rm kin}$-diagram (at
$\Theta=175$~\kms).
In contrast, the Bottlinger diagram and the histogram of normalized
$z$-extent are qualitatively the same as with the base-line sample.

Thus, we derived the scale heights for this restriced
sample via a $\chi^2$-fit from the $z$-probability distribution
results for the disks in $h_{\rm thin}=0.12$~kpc and $h_{\rm thick}=0.46$~kpc.
For the halo we get $h_{\rm halo}=4.2$~kpc.

\subsubsection{Non HB stars}
\label{bluered}

In the selection window there is a potential contamination of stars that do not
belong to the HB, such as massive core helium burners or RG stars
(see Sect.\,\ref{RHBRC}).

Core Helium burners are more massive than RHB stars and evolve faster.
The probability to find these inside our window is small.

RG stars of solar metallicity found within our selection window are younger
than 1.25 billion years (Yonsei-Yale isochrones by \citealt{YY}).
Assuming a constant Star Formation Rate (SFR) and an observed RGB
only of solar metallicity, we would
expect that an RGB of \mh$=-0.5$~dex would lie roughly 0.25~mag to the blue.
RG Stars of that metallicity develop faster by some 50\% (derived from
Yonsei-Yale evolutionary tracks by \citealt{Yi03}), so this metal-poor
RGB would in the CMD be at two-thirds of the observed RGB's density.
However, the density bluewards of the observed RGB is much lower than that.
Had the SFR in the past been smaller, then the contamination of our sample is
smaller, too.
If the SFR had been larger in the past, we would see many more RGs of bluer
colour than actually are seen in the \hip\ CMD.
All this shows that a contamination of our RHB sample (if present) comes from
young stars and not from old, metal-poor RG stars.

We check for this contamination of our RHB sample by separating the
selection window in two parts.
The Blue Sample is defined as having \bv$\leq0.85$~mag, while the other stars
are assigned to the Red Sample (see Fig.\,\ref{CMDsel}).
Stars that belong to the bluer sample are certain to be RHBs.
These 99 stars include 6 HPS stars.
The Red Sample contains 1228 stars, including 19 HPS stars.

The analyses of the two samples suggest that there are no
significant differences in kinematic behaviour.
The Blue Sample does have a comparatively higher fraction of halo stars,
including the two stars that have orbits reaching out to $z>10$~kpc.
Scale heights of the Blue Sample are found to be
$h_{\rm thin,blue}=0.1$~kpc,
$h_{\rm thick,blue}=0.8$~kpc, and $h_{\rm halo,blue}=13$~kpc.
For the Red Sample we obtain $h_{\rm thin,red}=0.12$~kpc,
$h_{\rm thick,red}=0.46$~kpc, and $h_{\rm halo,red}=1.8$~kpc.
The latter value is so low because the far-out moving halo stars are part of the
Blue Sample.

Had our base-line sample been
contaminated by stars of non-HB nature, in other words mostly younger stars,
then the $z$-distribution from the various samples should show significant
differences in the thin disk component.
This was not observed and we therefore conclude
that we are essentially working with just RHB stars.


\section{Effects of galactic mass models}
\label{models}

\begin{figure}
\centering
\epsfig{file=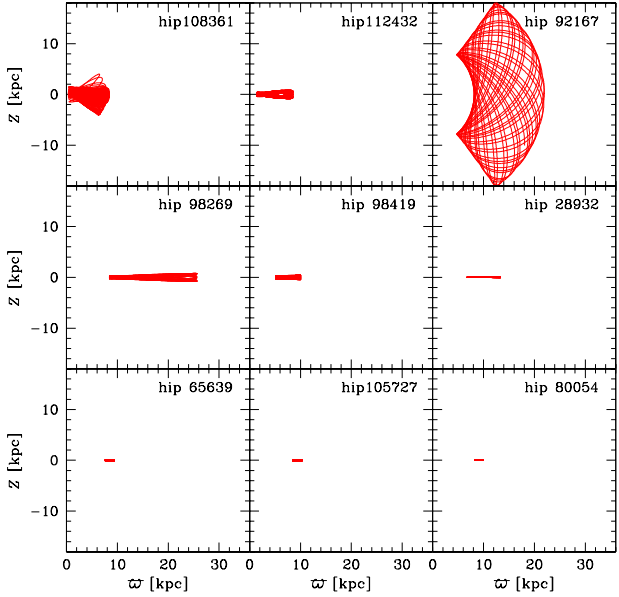,scale=0.43, angle=0}
\caption{Meridian plots of some of the calculated orbits in the ``2b''
potential by \cite{DB98}. The same stars as in Fig.\,\ref{orbs} were used to
allow comparisons.
The orbits shown are the ones differing most between the AS and DB potential
models (see Sect.\,\ref{models}).
\label{orbs_deh}}
\end{figure}

Several models have been proposed to describe the true gravitational potential
of our galaxy, e.g., by AS, \cite{Flynn96}, \cite{JSH95}, and
\cite{DB98} (hereafter DB).
Thusfar we have used the AS model, with $\varpi_{\odot}=8.5$~kpc and
$\Theta=220$~\kms.
Here we want to compare our results for the orbit statistics with those derived
using one of the DB models.
They are using different shapes for the disk and spheroidal potentials.
We traced the orbits of our selected RHB candidates in their
potential called ``2b''.
It was chosen, because with $\varpi_{\odot}=8.5$~kpc and $\Theta=231$~\kms\
it is the one most similar to the AS model.
Fits to the models yield the following Oort constants:
$A=12.95$~\kms\,kpc$^{-1}$ $B=-12.95$~\kms\,kpc$^{-1}$ for the model by AS
and $A=13.8$~\kms\,kpc$^{-1}$ $B=-13.3$~\kms\,kpc$^{-1}$ for the ``2b'' model by
DB.
To further facilitate comparisons with previous results, we chose the same time
constraints (10~Gyr backwards in 1~Myr steps).

\begin{figure}
\centering
\epsfig{file=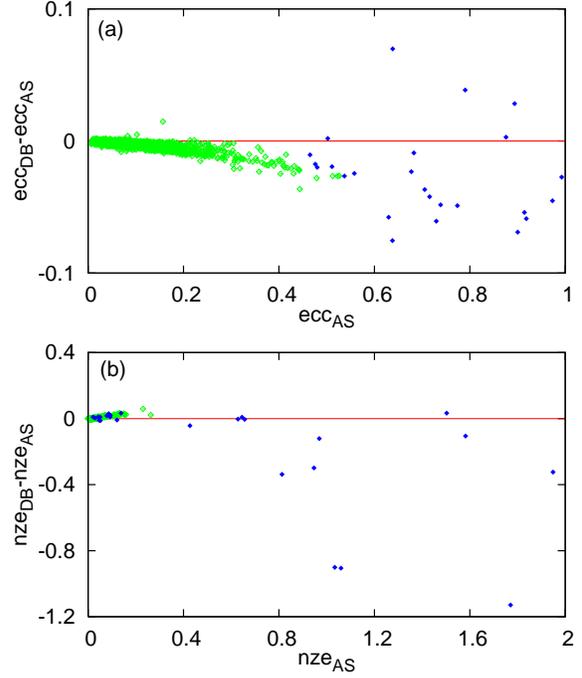,scale=0.45, angle=0}
\caption{Comparison of the eccentricities (a) and normalized $z$-extents (b)
of our stars in the potential of Allen \& Santillan (AS) and the ``2b'' potential of
Dehnen \& Binney (DB). The HPS stars are given as filled
symbols, the others as grey open circles. Orbits are less eccentric in the DB
than in the AS potential. In $nze$ there is a large scatter in the differences.
\label{EN_deh}}
\end{figure}

In Fig.\,\ref{orbs_deh} we provide the meridional plots of the same stars as in
Fig.\,\ref{orbs} from the ``2b'' potential.
The orbits of the first two objects shown appear to have changed considerably
in the meridional plot,
but they still retain almost the same eccentricities and normalized $z$-extents.
For the following two, reaching far out in $\varpi$, $ecc$ and $nze$
change.
Since these stars stay closer to the galactic center in the ``2b''
potential of DB, the $ecc$ decreases, while $nze$ increases.
Other orbits for disk stars appear to be largely unchanged on the plotted
scales.

Differences in $ecc$ and $nze$ between the orbits in the AS and DB potentials
for the whole 30\%\ ensemble are given in Fig.\,\ref{EN_deh}.
On average, a slight decrease of $ecc$ can be seen in Fig.\,\ref{EN_deh} (a),
while in normalised $z$-extent (Fig.\,\ref{EN_deh} (b)) the
disk stars lie slightly above zero, documenting larger values of $nze$ in the
``2b'' potential.
A closer look at the data revealed this to be a simultaneous effect of an
increase of $z_{\rm max}$ and decrease of $\varpi(z_{\rm max})$.
These differences arise from the different characteristics of the two
potentials.

\begin{figure}
\centering
\epsfig{file=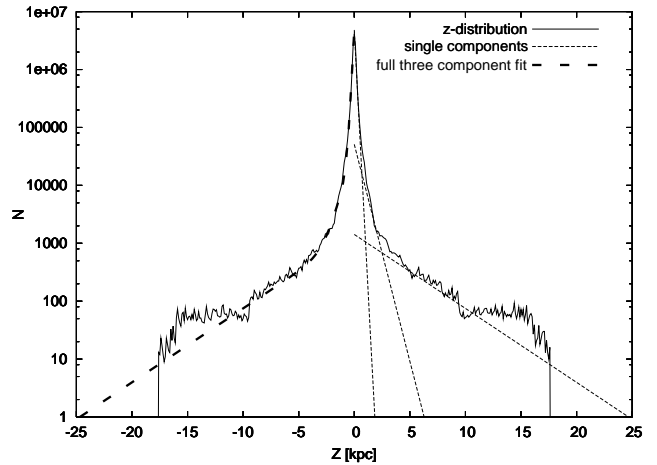,scale=0.35, angle=270}
\caption{The $z$-distribution of our \delpi$\leq30$\% RHB sample in the ``2b''
potential by DB.
The maximum $|Z|$ values are roughly 18~kpc, which is lower than in a potential
from AS.
This flatter distribution of the halo group is reflected in the fit
parameters:
$h_{\rm thin}=0.12$~kpc; $N_{\rm thin}(0)=4.8\cdot10^6$;
$h_{\rm thick}=0.58$~kpc; $N_{\rm thick}(0)=51000$;
$h_{\rm halo}=3.4$~kpc; $N_{\rm halo}(0)=1700$.
\label{Zhist_deh}}
\end{figure}

Medianised values of $\Theta$ and $E_{\rm kin}$ are found to be similar as in
Fig.\,\ref{ekin}.
Again, strong clustering along a roughly straight line is seen, which reaches
from the model's LSR rotational velocity of 231~\kms\ towards lower energies.
The gap in that line around $\Theta=115$~\kms\ is also seen, but it
is not as wide as for the AS data.

The $z$-probability distribution of our RHB candidate stars in the ``2b''
potential of DB is given in Fig.\,\ref{Zhist_deh}.
Differences to the fit parameter values for the AS potential only
arise for the halo.
Its lower scale height and higher number density are, however,
in line with the change in the shape of the orbits (compare Fig.\,\ref{orbs}
with Fig.\,\ref{orbs_deh}).


\section{Comparing the distribution of RHB stars with other types}
\label{final}

The results for the spatial distribution of the RHB stars can be compared with
the distributions found for other stellar types.
In particular a comparison with the distribution of sdBs (\citealt{AB2k4}) is of
relevance.

\subsection{General comparison}
\label{comp}

In Table\,\ref{fits} we have collected the data from our fit to
the $z$-probability distributions of the 30\% sample.
We give values for all three components, the thin disk, thick disk, and halo.
While the $N(0)$ values cannot be compared to real midplane densities, their
ratios can still be used for this discussion.

For the thin disk, \cite{Maj93} gives in his review a lower scale height limit
of 325~pc, a value mostly from successful starcount models.
Recently, \cite{Chen01} found $330\pm3$~pc, while \cite{Du03} found
$320\pm15$~pc.
Our value of 120~pc is much smaller than those.
However, there are measurements by \cite{Kui89b}, \cite{K93}, and \cite{Ojha94}
that give
249~pc, 250~pc, and 240~pc, respectively, as scale height for the thin disk.
It is not obvious why all these results differ so much.
Since we are using calculated orbits to determine the scale heights,
instead of just positions, the potential model may also come into play.

The review by \cite{Maj93} gives thick disk scale heights of 1.3 to 2.5~kpc.
Some more recently published values are 0.58-0.75~kpc (\citealt{Chen01}),
$0.64\pm0.03$~kpc (\citealt{Du03}), and 0.8-1.2~kpc (\citealt{Kerber01}).
Thus, the scale height of the thick disk is found to be in the range of
0.6 to 2.5~kpc.
Our value of 0.58~kpc for both potentials is at the lower end of the above
range.
Since the more recent studies tend to find values of 1~kpc or less,
we think that we are in line with those publications.

Midplane thick-to-thin-disk number ratios in the literature are given from as
low as 2\% (\citealt{Gil84}) up to 13\% (\citealt{Chen01}).
There is, due to the way such data are fitted, an anti-correlation between
density ratio and scale height.
Our ratio of about 1\% is lower than the values given above with a
rather low scale height.
The ratio of halo to thin disk stars taken from our data is
$N_{\rm halo}(0)/N_{\rm thin}(0)\simeq0.02$\% ($\simeq0.04$\% for the ``2b''
potential by \citealt{DB98}).
Typically, these values range from 0.05\% to 0.4\% (see e.g. \citealt{Chen01},
\citealt{Kerber01}).
These discrepancies are probably due to the small volume in space, from which
we drew our data, the small sphere around the Sun.
Other studies are based on selected fields with, in general, more distant
stars.
These are less likely to include thin disk stars, while thick disk stars are
less likely to be included in our sample.

Could this aspect (samples of different distance ranges) explain all the
differences?
A comparison between a spherical and conical survey of stellar
populations with the same densities and scale heights gives hints to the answer.
For a sphere the number of disk stars in the volume increases with the observed
distance.
This increase is much greater than for a narrow cone that looks away from the
disk.
For example, with our fit results from Sect.\,\ref{res3} we obtain a ratio of
$N_{\rm halo}/N_{\rm thin}\simeq0.17$\% for a sphere with radius 2~kpc, while
the same ratio for the cone of the same length towards a galactic pole is 5.5\%.
For a distance of 5~kpc the numbers are 0.33\% and 43\%, respectively.
This extreme case should emphasise, that even with our very low midplane density
ratio, a conical survey can find a large number of halo stars.
Effects arising from the shape and depth of the surveyed volume are therefore
important.

Another, more fundamental, reason for the different results comes
from the method of analysis.
While some other studies explicitly assign the stars to one of the three
stellar populations (thin disk, thick disk, halo), we opted not to do so.
Our results are not influenced by errors in such assignments.
Rather, our derived scale heights and number densities are only a result of the
fitting procedure and are dependent of the weights used.
Differences coming from the fit procedure are thus not negligible.
We chose inverse counts as weights, because the fit function approximated the
distribution equally well for both the disk and halo component.

\subsection{Are there differences in population along the HB?}
\label{HBpop}

In \cite{B97} the $z$-probability distribution was introduced as a tool to
determine scale heights of stellar populations.
The latest results on BHB stars involving this method are given in \cite{AB2k4}.
From some 120 sub-dwarf stars, they derived a scale
height of $0.9\pm0.1$~kpc for the thick disk.
Another population with scale height of about 7~kpc was found in the data.
These values are larger than the ones derived from our RHB sample (0.58~kpc and
4.2~kpc) by almost a factor of two.

For this paper the same orbit calculation tools were used as in \cite{AB2k4}.
Any systematic effects (too restrictive potentials as in Sect.\,\ref{comp})
arising from the analyses should therefore be the same for both studies.
But the fitting of the $z$-distributions was done differently.
We used the whole range in $Z$ to do a weighted fit for three
components simultaneously.

However, the values are taken from data of two positionally very different
samples.
The RHB sample is located near the sun, while
the sdBs are generally farther away at high galactic latitudes.
Consequently, from the RHBs we find $z$-distributions that lack a well defined
halo, while for the sdBs a thin disk component was barely seen.

The large difference for the halo scale heights from the RHBs results from this
``undersampling'' in our sample.
Another effect of this is that we get the lower midplane halo-to-disk
number density ratios mentioned in Sect.\,\ref{comp}.

A greater number of far-out halo stars could surely be found in the \hip\ and
BB2k data, but we deciced to limit the observed volume by our already
rather large relative parallax error of 30\%.
{\sc Gaia} (\citealt{Per01}), ESA's future astrometric mission will measure
three-dimensional positions and velocities down to 17.5~mag.
With such a wealth of data, we will be able to derive firmer results on the halo
scale height of RHBs.

The thick disk components from both studies are well represented, showing a
large number of objects.
Scale height differences in this population can therefore not be explained
by low number statistics.
Again, the selection of the samples influences the results, albeit at a lower
level than for the halo component.
Whether the scale heights can be explained by an intrinsic difference in
kinematics of RHBs compared to sdBs, cannot be decided.
For that we would need to observe the same volume of space.

\subsection{Summary of the results}
\noindent
{\bf 1)} 
The position and velocity diagrams (Figs.\,\ref{Toomre} and
\ref{bott}) show that the RHB sample 
is a mix of disk stars (with $\Theta \simeq 200$ \kms) 
and stars with lower orbital speed and thus larger perpendicular velocity, 
establishing the group of the asymmetric drift. 
A relatively small subset exhibits velocity and kinetic energy of a nature 
typical for the halo.

\noindent
{\bf 2)} 
The orbit shapes indicate a predominance of rather circular orbits.
Again, a small subset exhibits very elliptic orbits in part reaching to 
large distances from the disk.

\noindent
{\bf 3)} 
The orbit statistics make clear that only a few of the \hip-BB2k
stars reach far into the halo. 
This does not imply that stars found to have highly excentric orbits 
but staying in the disk are not part of the halo population. 

\noindent
{\bf 4)} 
No significant dependence of the thin disk parameters on the volume surveyed was
found.
However, the thick disk and halo show varying parameters due to the low number of
objects.

\noindent
{\bf 5)}
Different models of the distribution of mass in the galaxy do not
produce significant differences in the scale heights of our sample.

\noindent
{\bf 6)}
The vertical scale height of our RHB star samples and the ones of the earlier
investigated sdB stars are significantly different.
This either points at a different history of the progenitors of these groups
or, given the widely different spatial sampling, demonstrates that source
selection leads to strong biases with
concomittant differences in the resulting average parameters.

\noindent
{\bf 7)}
The range of thick disk scale heights, as derived from orbits as well as other
methods,
is either due to sample selection problems or is
inconsistent with a simple galactic structure.


\begin{acknowledgements}
We are grateful to the DLR for supporting TAK and MA under the grants
50QD 0103 and 50QD 0102, as well as FONDAP for support of MA under the grant
1501 0003.
Also, we wish to express our gratitude to Walter Dehnen, who gave us access to
his galactic mass models and software for orbit calculation.
We like to thank  Michael Odenkirchen and Michael Geffert for
providing program code used throughout these studies, and Philip Willemsen
for helpful discussions.
For this work the SIMBAD and VIZIER databases at CDS in Strasbourg were
frequently used.
\end{acknowledgements}

\bibliographystyle{aa}
\bibliography{tkaempf}

\end{document}